\documentclass{sig-alternate}

\usepackage{amsthm}
\usepackage{amssymb}

\newtheorem{theorem}{Theorem}
\newtheorem*{theorem2}{Theorem}

\newtheorem{prop}[theorem]{Proposition}
\newtheorem*{prop2}{Proposition}
\newtheorem{defn}[theorem]{Definition}
\newtheorem{algorithm}[theorem]{Algorithm}
\newtheorem{example}[theorem]{Example}
\newtheorem{cor}[theorem]{Corolary}

\DeclareMathOperator{\signum}{\mathtt{signum}}
\DeclareMathOperator{\floor}{\mathtt{floor}}
\DeclareMathOperator{\expand}{\mathtt{expand}}
\DeclareMathOperator{\normal}{\mathtt{normal}}

\DeclareMathOperator{\fin}{\mathtt{Fin}}
\DeclareMathOperator{\cfin}{\mathtt{\mathbf{Fin}}}
\DeclareMathOperator{\Hom}{\mathtt{Hom}}

\DeclareMathOperator{\pnform}{\mathtt{pseudonormalform}}
\DeclareMathOperator{\cform}{\mathtt{canonform}}

\newcommand{\undefined}{\bot}
\newcommand{\reals}{\mathbb R}
\newcommand{\creals}{\overline{\reals}}
\newcommand{\ZZ}{\mathbb Z}
\newcommand{\Q}{\mathbb Q}
\newcommand{\RP}{\mathcal R}
\newcommand{\Chi}{\mathcal X}

\begin{document}

\setcounter{topnumber}{3}
\setcounter{dbltopnumber}{3}
\setcounter{totalnumber}{6}
\setlength{\floatsep}{4pt plus 2pt minus 3pt}
\setlength{\textfloatsep}{4pt plus 2pt minus 3pt}
\setlength{\intextsep}{4pt plus 2pt minus 3pt}
\setlength{\dblfloatsep}{4pt plus 2pt minus 3pt}
\setlength{\dbltextfloatsep}{4pt plus 2pt minus 3pt}
\setlength\itemsep{0pt plus 1pt minus 1pt}
\renewcommand\floatpagefraction{.95}
\renewcommand\topfraction{.95}
\renewcommand\bottomfraction{.95}

\title{A canonical form for some piecewise defined functions}

\numberofauthors{1}
\author{
Jacques Carette\titlenote{Supported in part by NSERC Grant Discovery Grant RPG262084-03}\\
       \affaddr{McMaster University}\\
       \affaddr{1280 Main Street West}\\
       \affaddr{Hamilton, Ontario, Canada}\\
       \email{carette@mcmaster.ca}
}

\maketitle
\begin{abstract}
We define a canonical form for piecewise defined functions.
We show that this has a wider range of application as well as better
complexity properties than previous work.
\end{abstract}


\section{Introduction}\label{intro}
Piecewise defined functions are ubiquitous in mathematics, starting from
the Kronecker Delta function, through characteristic functions for sets,
on to functions such as $\signum$ and $\floor$.  Although all of these are
certainly interesting, this paper will concentrate on those functions
defined over a linearly ordered domain (like $\reals$ or $\creals$) and with a
finite number of pieces (unlike $\floor$ say).  

There has been previous work in this area, most notably that of
von Mohresnchildt \cite{Mohrenschildt}.  There, a
normal form was defined for a large class of piecewise-defined expressions
through the use of a very simple set of primitive elements; as well, clear
steps were given to modify this normal form to give a canonical form.  In our
approach, the primitive elements are much more complex; however this allows all
the algorithms to be greatly simplified.  Furthermore,
we obtain substantial arithmetic complexity improvements.
We can also handle a wider domain of definition.
This form has been independently rediscovered by
several authors, see for example \cite{MohBau} and \cite{Borwein}.
However both of those papers are about applications of these extended
piecewise functions to optimisation, to Fenchel coordinates in particular.
To our knowledge, the underlying theory of piecewise functions over
linearly ordered spaces has never been published.

It is important to note that, outside of \cite{Mohrenschildt} (and the
references therein), there seems to
be no reference to a formalization of the concept of a piecewise
function.  This is probably because the usual notation is so suggestive that no
one ever thought to question if the concept was ever properly defined.

The results we obtain in this work are deceptively simple, but this is
largely because a considerable amount of effort has been put into ensuring
that all the definitions are ``just right''.  

This paper benefited from some discussions of the contents with Alexander
Potapchik of Maplesoft Inc.  He also implemented, in Maple 7, many of the ideas
contained in this paper, and this is what Maple now uses for simplification and
normalization of piecewise functions.

\section{Piecewise}\label{canon}

\subsection{Observations}
Although the most common piecewise defined functions are of the
type
\begin{equation}\label{E:pw1}
f(x) = \begin{cases}
-1 & x<0\\
1 & \text{otherwise.} 
\end{cases}
\end{equation}
where $x$ is (implicitly) understood to be real, we also encounter functions
of the kind
\begin{equation}\label{E:pw2}
 f(x) = \begin{cases} 
x^2 & y<0\\
x^3 & \text{otherwise}
\end{cases}
\end{equation}
where $x$ and $y$ are also (implicitly) understood to be real.  The notation
in the second case above is poor, as the dependence on $y$ is not well
indicated, but in frequent use nevertheless.  This leads us to observe that, in
both cases, there really are two different kinds of variables at play: those
that need to satisfy a boolean condition, and those that occur in an arithmetic
context ($y$ and $x$ respectively).  This ``separation of 
concerns'' leads to an important conceptual simplification of the requirements
for a piecewise defined function.

Another observation is that, at least in computer algebra systems, it is
common to take the derivative of objects like
$$ f(x) = \begin{cases}
-1 & x<0\\
1 & \text{otherwise.} 
\end{cases}
$$
Accordingly, the resulting object
$$ f'(x) = \begin{cases}
0 & x\neq 0\\
\undefined & \text{otherwise} 
\end{cases}
$$
should really be within the realm of objects that can be talked about.
Furthermore it should be possible to correctly compute with such partial
functions, as well as with functions on extended domains.  The normal
form of \cite{Mohrenschildt} explicitly requires a ring for the range.

A third and final observation is that, for linearly ordered domains like 
$\reals$, adding and even multiplying two functions that are each defined by
formulas valid on some finite union of intervals is very easy and can be done
in linear arithmetic cost (with respect to the total number of intervals).
From these three observations, the author believes that a keen reader
should be able to derive the rest of the paper!

\subsection{Definition of piecewise}\label{defn:piecewise}

We will start with a relatively simple case of piecewise-defined function, one
which is defined on a unique linearly ordered domain.

\begin{defn}
A set $S$ is said to be \textbf{linearly ordered} if there exists a relation $<$
on $S$ such that for all $a,b\in S$, $a\neq b$ either $a<b$ or $b<a$ holds.
\end{defn}
From now on, let $\Lambda$ be a linearly ordered set.  We will also need the
concept of \textbf{range partition} of such a set.  This is one of the crucial 
ingredients.

\begin{defn}
A \textbf{range partition} $\RP$ of a linearly ordered set $\Lambda$ is a finite set 
$B$ of points $\lambda_1<\lambda_2<\ldots<\lambda_n$, along with the natural
decomposition of $\Lambda$ into disjoint subsets subsets 
$\Lambda_1, \ldots, \Lambda_{n+1}$ where
$$\Lambda_1 := \{x\in \Lambda~|~x<\lambda_1\}$$
$$\Lambda_i := \{x\in \Lambda~|~\lambda_{i-1}<x<\lambda_i\},i=2,\ldots,n$$
$$\Lambda_{n+1} := \{x\in \Lambda~|~\lambda_n<x\}.$$
\end{defn}
\noindent
Note that the $\lambda_i$ themselves are outside these subsets, that
$$\Lambda = \left(\bigcup_{i=1}^{n+1}\Lambda_i\right) \cup 
            \{\lambda_1,\ldots,\lambda_n\}
          = \bigcup\RP,$$
and that it is the {\em ordered} version of this complete decomposition of
$\Lambda$ which is the range partition.
For a given $\Lambda$, we will often just give the set of points $\lambda_i$ 
that generate a range
partition.  It is sometimes useful to consider $\Lambda$ itself to be a
degenerate range partition with the empty set $\emptyset$ as the generating set.
We will sometimes refer to the generating set $B$ of a range partition as a set
of \textbf{breakpoints}.

Perhaps surprisingly, it is expressions like (\ref{E:pw2}) that are simplest
to deal with.

\begin{defn}
A \textbf{piecewise expression} is a function from a range partition to a set
$S$.
\end{defn}

\begin{example}
Taking $\Lambda=\reals$, $B=\{0\}$, and 
$S=\{x^2, x^3\}$ then $f:\RP\to S$ defined by
$$f(z) = \begin{cases} 
x^2 & z=\Lambda_1\\
x^3 & z=0\\
x^3 & z=\Lambda_2,
\end{cases}
$$
is a piecewise expression.
\end{example}
Of course this is a rather pedantic definition as this clearly does not
represent an object of common mathematical interest.  Nevertheless it is a very
useful definition as it encodes the core computational concept necessary for
the sequel succinctly and unambiguously.  With just a little
more work, we will soon be able to define an object which will be much closer
to the usual piecewise functions encountered in textbooks.

\begin{prop}
Let $\Lambda$ be a linearly ordered set and $\RP$ a range partition.  Then there
exists a function $\Chi:\Lambda\to\RP$ which associates to each $\lambda\in\Lambda$
the unique element $r$ of $\RP$ such that $\lambda\in r$.
\end{prop}

\begin{cor}\label{linear_time}
Assuming that $=$ and $<$ are decidable and 
take $O(1)$ time, then for $\lambda\in\Lambda$, $\Chi(\lambda)$ 
can be computed using at most $O(log_2(n))$ operations, where $n=|B|$.
\end{cor}

\begin{proof}
Since $\Lambda$ is linearly ordered, we can store $\RP$ in a contiguous sorted 
array and use an adapted binary search on its $2n+1$ elements to find 
$\Chi(\lambda)$.
\end{proof}

The assumption that $=$ and $<$ are decidable over all of $\Lambda$ can
be weakened to merely assuming that they are decidable for the evaluation
point $\lambda$ {\em relative} to be set of breakpoints $B$.  This is why in 
practice these functions can be effectively evaluated even though 
the zero equivalence problem is undecidable.

From now on we will assume that all range partitions are stored in a contiguous
sorted 1-dimensional array; we will sometimes simply say use the term 'list' to
refer to this data-structure.

Using $\Chi$, and a little bit of abusive notation, we get a much more familiar
expression for $f_B=f\circ\Chi:\Lambda\to S$ where we explicitly indicate the
range partition generator $B$.  For the previous example, this unravels to:
$$f_{\{0\}}(z) = \begin{cases} 
x^2 & z<0\\
x^3 & z=0\\
x^3 & z>0.
\end{cases}
$$
There is clearly a bijection between the set of $f_B$ and the set of piecewise
expression defined previously.  Next, we really want to be able to treat
expressions like
\begin{equation}
\label{abs} f(x) = \begin{cases}
-x & x<0\\
0 & x=0\\
x & x>0.
\end{cases}
\end{equation}
where want the evaluation bindings to be such that $f(-5)=5$ and not $-x$.
Our definition of piecewise expressions, using terms from a set $S$ as
above, would indeed give $-x$ because there is no relationship between
the elements of $\Lambda$ and those of $S$.  This is definitely not what is
wanted.  If we used expressions with strict evaluation rules, this particular
problem would be solved.  However, that is not quite enough because we would
still have problems with singular expressions in ``other'' branches.
To fix both of these problems at
once, what we really need to do is to treat $S$ as a set of functions instead
of a set of (first order) values.  To avoid spurious evaluations, we
are going to steal a standard trick\footnote{also known to logicians as
lambda-lifting} from functional programming\footnote{we could have also used
some fancy version of lazy evaluation, but that would have introduced new
problems whose solution would have distracted greatly from the main points of
this paper.  A very specialized version of lazy evaluation is what was later
implemented in Maple 8 for this purpose but, in this author's opinion, the
downsides of integrating this in an eager language outweigh the apparent
benefits of being able to use a simpler representation.} and use currying to
solve our problems.  This leads us to define a somewhat more general concept
than a piecewise function, but the extra generality is exactly what allows us
to solve the problem mentionned above.  Furthermore, it specializes easily and
correctly to the intuitive notions of piecewise functions, as we will prove in
the next section.

\begin{defn}
Let $S$ be a set of functions, then a \textbf{piecewise operator} is a 
piecewise expression $f:\RP\to S$.
\end{defn}
We can thus rewrite example~\ref{abs}, using $\tilde{S}=
\{y\mapsto -y, y\mapsto 0, y\mapsto y\}$, the curried, relabelled version of 
$S$ to get 
\begin{equation}\label{curried_abs} \tilde{f}(x) = 
\begin{cases}
y\mapsto -y & x<0\\
y\mapsto 0 & x=0\\
y\mapsto y & x>0.
\end{cases}
\end{equation}
Then we have that $\tilde{f}(-5)(\sqrt{2})=-\sqrt{2}$.  This is actually
progress!  What we really want is $\tilde{f}(-5)(-5) = 5$.  This 
last ingredient is exactly what we need to define piecewise functions that
behave as expected mathematically as well as when implemented.

\begin{defn}
Given a piecewise operator $f:\RP\to S$ where $S$ is a set of functions
${s:\Lambda\to V}$ call $\overline{f}:\Lambda\to V$ defined by
$$\overline{f}(\lambda) := f(\Chi(\lambda))(\lambda) = f_B(\lambda)(\lambda)$$
a \textbf{piecewise function}.
\end{defn}
Note that the notation $\overline{f}$ is sufficient since all of $\RP, \Chi$ and $B$ 
can be recovered from a {\em representation} of $f$.  Also note that there are
no restrictions on $V$ at all.  When multiple piecewise functions
defined on different range partitions (but the same $\Lambda$)
are being discussed, we will denote them $p_B$, making the generating set of the
range partition explicit.  A strict notation for piecewise functions would be
given by
$$
f(x) := \begin{cases}
g_1(x) & x\in\Lambda_1\\
g_2(x) & x=\lambda_1\\
g_3(x) & x\in\Lambda_2\\
g_4(x) & x=\lambda_2\\
\vdots & \vdots\\
g_{2n}(x) & x=\lambda_n\\
g_{2n+1}(x) & x\in\Lambda_{n+1}
\end{cases}
$$
with $g_i\in S$.  It is worthwhile noting that giving $\Lambda, B$ and 
$g_1,\ldots,g_{2n+1}$ (as ordered sets) are sufficient to fully determine $f$.

As $\Lambda$ is linearly ordered, and the $g_i$'s for $i$ even are actually 
only evaluated at one point, this is customarily written as
$$
f(x) := \begin{cases}
h_1(x) & x<\lambda_1\\
\beta_1 & x=\lambda_1\\
\vdots &  \vdots\\
\beta_n & x=\lambda_n\\
h_{n+1}(x) & \lambda_n<x
\end{cases}
$$
where that last condition is often written as the word 
\textbf{otherwise}, and where $h_i=g_{2i-1}$ and $\beta_i=g_{2i}(\lambda_i)$.
This notation can at times be problematic as it mixes ground values (the 
$\beta_i$'s) and functions (the $h_i$) at the same ``level'', even though
they have different types.  This is why we prefer to lift the constants up 
to functions explicitly.

It is important to notice that a piecewise function is a function that uses its
argument twice, for very different purposes.  It is the separation of these two
concerns that make many of the subsequent algorithms simple yet general.  In
the next sections it will be important to keep track of which properties hold
in the general case of piecewise operators and which need to be specialized for
piecewise functions.

\subsection{Definition of domains}

In order to be able to define a canonical form, we will require somewhat more 
structure on the range $S$ of functions of a piecewise operator. 

\begin{defn}
An \textbf{effective domain} $D$ is a pair $(F, \sim)$, where
\begin{enumerate}
\setlength\itemsep{0pt plus 1pt minus 1pt}
\item $F:O^n\to V$ is a set of functions (of varied arity $n$) from a set $O$ 
to a set $V$
\item $\sim$ is a binary function on $F$ that decides extensional equivalence.
\end{enumerate}
\end{defn}

\begin{defn}
Two (n-ary) functions $f,g\in F$ are said to be 
\textbf{extensionally equivalent} if for all $x\in O^n$, either $f$ and $g$ are
both defined and $f(x) = g(x)$, or neither $f$ nor $g$ are defined.  Denote this
by $f \simeq g$.
\end{defn}

\noindent It is very important to note that 
\begin{enumerate}
\setlength\itemsep{0pt plus 1pt minus 1pt}
\setlength\parsep{0pt plus 1pt minus 1pt}
\item the functions in $F$ can be partial,
\item $\simeq$ denotes equivalence, not equality,
\item $\sim$ is defined for $F$, not $O$ nor $V$,
\item $\sim$ decides equivalence, where $\simeq$ denotes equivalence.
\end{enumerate}

In most practical cases, $\sim$ will necessarily be compatible with a
(possibly partial) equivalence of elements of $V$ since
there is a canonical identification between the functions $g_{2i}$ of
the breakpoints of a piecewise operator and the constants they represent.  
Of course, since $\sim$ is a decision procedure, this implies that the
constant functions present in $F$ must in fact come from a subset of
constants of $V$ over which a similar decision procedure exists.
But the point is 
that we should be able to tell that $(x+1)^2$ and $x^2+2x+1$ (over $\reals$ say)
are equivalent.  What is crucial here is that we can tell this completely
independently from any representation issue of the underlying domain.  In 
other words, this works just as well over the usual uncountable $\reals$ as
it does with constructive $\reals$.

Given an effective domain $D$ and a computable total function $C:F\rightarrow
F$ such that $C(s)\sim s$, $C(s)=0 \iff s\simeq 0$, and
$\forall s,t\in F,\ C(s)=C(t) \iff s\sim t$, we will call the
triple $\{F, \sim, C\}$ a \textbf{strong effective domain}.
It is worthwhile
noting that given $\{F, C\}$ one can always obtain a strong effective domain
by defining $\sim$ to be $(a,b)\mapsto C(a)=C(b)$ whenever equality is decidable on
$F$.  

\begin{prop}\label{canonical}
Let $\{F, \sim, C\}$ be a strong effective domain.  Then $C\circ C = C$.
In other words, $C$ is a canonical form for $F$.
\end{prop}

\begin{proof}
Let $s\in F$, and $t=C(s)$.  Since $t=C(s)\sim s$, then 
$C(t) = C(C(s))\sim C(s) \sim s$.  But $t=C(s)\sim s$, so $C(C(s)) = C(t) = C(s)$.
\end{proof}

One cannot under-estimate the power of such a $C$: it gives a {\em canonical form}
for functions in $F$.  It is important to notice that it is defined {\em globally},
in other words, it treats (partial) functions of the whole domain.  It is outside
of the scope of the current work, but roughly speaking such canonical forms only 
(seem to?) exist for very rigid objects, like meromorphic functions or, more generally
for functions for an $o$-minimal structure.

\begin{example}
The polynomial functions over $\ZZ$, coded as 
$D = \{ \ZZ[\bar{x}],\ =,\  \expand \}$, is a strong effective domain.
\end{example}

In fact, we can replace $\ZZ$ with $\mathcal{RA}$, the real algebraic numbers, and
still get a strong effective domain, see \cite{Kung:1973:CCA} for
the details.  This example also shows why it is important to deal with
equivalence rather than equality, as well as the fact that a canonical form
induces a (computable) equivalence test.  Perhaps more important still, at
least to symbolic computation, is the next example.
\begin{example}
Let $P$ be a term algebra (of rational functions)
containing the rationals $\Q$, the symbol $x$, 
the binary operations $+, \times$, and composition.  Let $T$ be the term
algebra defined by the grammar $P | \sin(P) | \cos(P) | T+T | T*T$.
Let $T'=\{f:\reals\to\reals~\text{where}~f:x\mapsto t~, t\in T\}$ 
be the corresponding set of functions.  Then $\{T, \normal\}$ with $\normal$
the expanded normal form defined in \cite{Monagan}, is a strong effective
domain.
\end{example}
While it is possible to further generalize the above example, 
the term algebra $T$ is already very close to the one used in the
undecidability results of \cite{caviness:simplification,Richardson:1968}, and
thus we cannot expect to be able to continue with pure decision procedures much
further, although it would be interesting to see in which ways holonomic
functions can be mixed with piecewise functions and retain decidability.
Semi-decision procedures and even heuristics can however be quite useful in
practice.

A \textbf{weakly effective domain} is a pair $\{F, \sim\}$ where
$\sim$ only decides equivalence to a distinguished element of $F$ (typically
$x \mapsto 0$).  This is often the case when associated to $F$ we have a
normal form operator $N$ for elements of $F$ instead of a canonical form.

\subsection{Spaces of piecewise operators}

\begin{defn}
Let $S$ be a set, then denote by $\fin(S)$ the set 
$\{p\in\mathcal{P}(S)~|~\sharp p<\infty\}$ of finite subsets of $S$, 
where we denote the power set of $S$ by $\mathcal{P}(S)$.
\end{defn}

\begin{defn}
Let $\mathbb{P}(\fin(\Lambda), F)$ denote the set of all piecewise operators defined
on the range partitions of $\Lambda$ generated by all its (finite) subsets with 
values in $F$.
\end{defn}

\noindent Of particular interest will be the case where $\{F,\sim\}$ is (at
least) a weakly effective domain.
As we will often discuss multiple piecewise functions at once, it is convenient to
define $B:\mathbb{P}(\fin(\Lambda), F)\rightarrow \fin(\Lambda)$ as the function which
given a (representation of a) piecewise operator will return its set of breakpoints.

\subsection{Redundancies and refinement}

It is important to notice that $F$ is canonically embedded in
$\mathbb{P}(\fin(\Lambda),F)$ since $\emptyset\in\fin(\Lambda)$ generates
piecewise operators extensionally equivalent to those in $F$.  However, this
space also contains a lot of redundancies.  If we let $\Lambda=\reals$ and $F$
the space of all continuous functions $C(\reals, \reals)$, then
$$p(\lambda) := \begin{cases}
1 & \lambda<0 \\
1 & \lambda=0 \\
1 & \lambda>0
\end{cases}
$$
is clearly an element of $\mathbb{P}(\fin(\reals), C(\reals, \reals))$ which is
extensionally equivalent to $x\mapsto 1\in F$.  We will deal with this
redundancy later.  This redundancy is in fact very useful, and is the key to
efficient arithmetic in $\mathbb{P}$! As increasing the redundancy of the
representation of a piecewise operator can be quite useful, we will encode this
in a definition.

\begin{defn}
A \textbf{refinement} of a piecewise operator $p$ is another operator $q$ such that
$p(\lambda)=q(\lambda)$ for all $\lambda\in\Lambda$, and
the set of breakpoints of $p$ is a subset of that of $q$.
We will call a refinement \textbf{strict} if the set of breakpoints of $p$ is a
strict subset of that of $q$.
\end{defn}
\noindent
Note that we used $p=q$ and not $p\sim q$ in this definition.  It is in fact
possible to do this either way, but since we will always be using explicit
refinements, this would be an unnecessary complication.  Most often, we will
actually want to specify the (new) set of breakpoints of a refinement:
\begin{defn}
For any ordered finite set $A\subset\Lambda$, 
a $A$-\textbf{refinement} of a piecewise operator $p$ is another operator $q$
such that $q$ is a refinement of $p$, and $A\subset B(q)$.
We will say a $A$-refinement $q$ is \textbf{exact} if $B(q) = A \cup B(p)$.
\end{defn}
Given a finite ordered set $A$ and a piecewise operator $p$ as above, one can
use the usual \textbf{linear merge} algorithm to generate $q$ in time $O(|A\cup
B(p)|)$.

We have glossed over one very important point: we can perform a linear merge of
two ordered finite lists of breakpoints if and only if we can effectively
decide $<$ and $=$ for each of the breakpoints.  In other words, for all of our
algorithms we need to make the assumption that whenever we need to compute a
common refinement $q$ of two piecewise functions $p_1, p_2$, then the union $B$
of their respective sets of breakpoints $B_1, B_2$ must be such that
$B\subset\cal{O}\subset$$\Lambda $ where $<$ and $=$ are decidable on $\cal{O}$.
For this purpose, we introduce following variation on $\fin$.  

\begin{defn}
Let $D$ be a linearly ordered domain, $\mathcal{O}$ a subset of $D$ over which
$<$ and $=$ are decidable, then denote by $\cfin(S)$ the set
$\{p\in\mathcal{P}(\mathcal{O})~|~\sharp p<\infty\}$ of finite subsets of
$\mathcal{O}$. 
\end{defn}

All definitions of piecewise functions, piecewise operators and operations on
them should be understood to use $\cfin$ in place of $\fin$ whenever
computability and decidability are needed.  We will not systematically do so
since the \textit{mathematical} definitions of many of the concepts work equally well
without this restriction.

Another aspect to notice is that since we are dealing with piecewise operators
(and not functions) even at breakpoints, so that the underlying functions in the
representation of $p$ are not evaluated to give $q$.  For example, the
$\{1\}$-refinement of the trivial piecewise operator 
$p(\reals) := (x\mapsto 0)$ is
$$q(\lambda) = \begin{cases}
x\mapsto 0 & \lambda<1 \\
x\mapsto 0 & \lambda=1 \\
x\mapsto 0 & \lambda>1.
\end{cases}
$$
\subsection{Denesting}\label{denest}

There are two different ways in which ``nesting'' of piecewise expressions
can arise: functional composition and definitional nesting.  This is easiest
to understand via examples:  consider the piecewise operators
\begin{equation}\label{funky}
t(\lambda) = 
\begin{cases}
x\mapsto x^2-3             & \lambda<1 \\
x\mapsto -5                & \lambda=1 \\
x\mapsto x^3-7x^2+16x-12 & \lambda>1.
\end{cases}
\end{equation}

\newcommand{\ovt}{\overline{t}(\lambda)}
\newcommand{\sovt}{\begin{cases} \lambda^2-3 & \lambda<1 \\ -5 & \lambda=1 \\ \lambda^3-7\lambda^2+16\lambda-12 & \lambda>1 \end{cases}}

\noindent and the absolute value function as the piecewise operator $f$ of 
example~\ref{curried_abs}, along with the corresponding piecewise functions 
$\overline{f}, \overline{t}$ .  Then $\overline{f}(\ovt)= |\ovt|$ is 
an example of functional composition.  Expanding the definitions gives
\begin{align*}
\label{funky_composition}
\overline{f}(\ovt) &= \begin{cases}
-\ovt & \ovt <0 \\
0     & \ovt =0 \\
\ovt  & \ovt > 0
\end{cases} \\
& = 
\begin{cases}
-\ovt & \left(\sovt\right) <0 \\
0     & \left(\sovt\right) =0 \\
\ovt  & \left(\sovt\right) > 0
\end{cases}
\end{align*}
\noindent which, after quite a number of non-trivial computations 
(see \cite{Mohrenschildt} for the details) gives
\begin{align*}
\overline{f}(\ovt) &= \begin{cases}
\lambda^{2}-3                          & \lambda< -\sqrt {3} \\
0                                      & \lambda= -\sqrt{3} \\
-\lambda^{2}+3                         & \lambda<1\\
5                                      & \lambda=1 \\
-\lambda^{3}+7\lambda^{2}-16\lambda+12 & \lambda<3\\
\lambda^{3}-7\lambda^{2}+16\lambda-12  & 3\leq \lambda
\end{cases}
\end{align*}
\noindent were we would have to expand the first and last cases further
if we wanted to write this more formally. The most difficult
parts of this computation involve extracting a range partition from
conditions like
$$\left(\sovt\right) <0$$
The case of definitional nesting is considerably simpler.
$$\begin{cases}
\left(
\begin{cases}
\lambda^2-3             & \lambda<1 \\
-5                & \lambda=1 \\
\lambda^3-7\lambda^2+16\lambda-12 & \lambda>1.
\end{cases}
\right) & \lambda < 3 \\
3 & \lambda = 3 \\
\left(
\begin{cases}
-\lambda   & \lambda < 0 \\
0         & \lambda = 0 \\
\lambda   & \lambda > 0
\end{cases}
\right) & \lambda>1
\end{cases} $$
\noindent only involves simple set-theoretic intersections to obtain
the equivalent
$$
\begin{cases}
\lambda^2-3             & \lambda<1 \\
-5                & \lambda=1 \\
\lambda^3-7\lambda^2+16\lambda-12 & \lambda<3\\
3 & \lambda = 3 \\
\lambda   & \lambda > 3
\end{cases}$$

\section{Arithmetic}\label{arith}

We first show how to do arithmetic with piecewise functions.  Very few
assumptions are needed to just perform arithmetic.  
For this section, let $\Lambda$ be a fixed linearly ordered set, and $F$ a set
of functions from $\Lambda$ to some set $M$.  Let
$\mathbb{P}=\mathbb{P}(\fin(\Lambda), F)$ be the corresponding space of
piecewise operators.  Furthermore suppose we have a function $\psi:F\to F$, we
want to lift this to a function on $\mathbb{P}$.

\begin{defn}
Let $\psi:F\to F$ be a unary function on $F$. For $p\in\mathbb{P}$ determined
by a breakpoint set $B$ and functions $g_i$, $1\leq i\leq 2|B|+1$, define
$\overline{\psi}(p)$ by the same breakpoint set $B$ and $\psi(g_i)$, $1\leq
i\leq 2|B|+1$.
\end{defn}

We should prove that this properly lifts the unary functions from those of $F$
onto $\mathbb{P}$:

\begin{theorem}
$\overline{\psi}(p)$ and $\lambda\mapsto\psi(p(\lambda))$ are extensionally
equivalent.  
\end{theorem}

\begin{proof}
Let $\lambda\in\Lambda_i$.  Then 
$$
\begin{aligned}
\overline{\psi}(p)(\lambda) &= \psi(g_{2i+1}) & & \quad \text{(by definition})\\
 &= \psi(p(\lambda)) & & \quad \text{(by definition)}
\end{aligned}
$$
as required.  Similarly, let $\lambda=\lambda_j$,
then $\overline{\psi}(p)(\lambda) = \psi(g_{2i}) = \psi(p(\lambda))$. 
\end{proof}

\noindent We can lift any n-ary function on $O$ to one on $\mathbb{P}^n$.  For
the case of $\psi:F\times F\to F$ and $p^1, p^2$ with common breakpoint set $B$
and associated functions $g^l_i$ for $l=1,2$, $1\leq i\leq 2|B|+1$,
$\overline\psi(p^1,p^2)$ is defined by the same breakpoint set and
$\psi(g^1_i,g^2_i)$.  The details for this and the n-ary case are left to the
reader - they are not difficult, but notationally hideous, and no new insight
is gained from the exercise.

For the case $\psi:F\times F\to F$ and $p^1, p^2$ with different breakpoint
sets $B_1, B_2$, we must first transform $p_1, p_2$ to $B$-refinements $q_1,
q_2$ (with $B=B_1\cup B_2$), and then we can apply the previous construction.
However we can no longer work over $\fin(\Lambda)$ but must work only over
$\cfin(\Lambda)$ for the refinement algorithm to be effective.

\begin{cor}
Addition from a linearly ordered ring $R$ can be lifted to addition of
piecewise-defined polynomials.  More generally, any ring $R$ gives rise to well
defined operations on $\mathbb{P}(\fin(|R|), R[x])$ with $|R|$ the underlying
set of elements of the ring $R$.
\end{cor}

\begin{proof}
The lifting of the addition from any ring $R$ to addition on $R[x]$ is
classical.  Treating $R[x]$ as a set of functions, one can use the previous
construction to lift addition (from $+$ to $\overline{+}$) up to
$\mathbb{P}(\fin(|R|), R[x])$.
\end{proof}

\noindent
Clearly the same can be done for negation and multiplication, and so on.

\begin{cor}
Let $R$ be a linearly ordered ring, and denote
$\Hom(R,R)$ the space of homomorphisms from $R$ to $R$.  Then
we can make $\mathbb{P}(\fin(R),\Hom(R,R))$ into a ring.
\end{cor}

\begin{proof}
Lifting the ring operations from $R$ to $\Hom(R,R)$ is classical: 
$(f+g)(x)$ is defined to be $f(x)+g(x)$, etc.  The functions
$\mathbf{0} = x\mapsto 0$ and $\mathbf{1}_x = x\mapsto x$ are the
additive and multiplicative unit respectively.  Letting $F=\Hom(R,R)$, 
simple verification shows that $(\mathbb{P}(\fin(|R|),F), \mathbf{0},
\mathbf{1}_x, \overline{+}, \overline{*})$ is a ring.
\end{proof}

\noindent
The previous corollary hints at an even more general result: that our
construction is actually functorial.  We will not go into the details since
this is not needed.
Before we move on, it is useful to explicitly turn these theoretical results in
to algorithms.  For example, using binary operators, we get

\begin{prop}\label{parith}
Let $R=\mathbb{P}(\fin(|O|,F)$ be a ring a piecewise functions.  Let
$f^1,f^2\in R$ be given explicitly.  Then $f^+=f^1 + f^2$ can be computed
explicitly by
\begin{enumerate}
    \item Forming the $B$-refinements of $f^1$ and $f^2$, with $B=B(f^1)\cup B(f^2)$.
    \item Letting the component functions $g_i^+$ of $f^+$ be
        $g_i^1 + g_i^2$, where the $g_i^j$ come from the $B$-refinements above.
\end{enumerate}
\end{prop}

\noindent 
Clearly we can replace $+$ by any other binary operation.  The reader will recognize
the above as being the \emph{linear merge} algorithm.

\section{Canonical form}

Simply defining arithmetic is not the end of the story.  For example, consider
$|x|^2-x^2$ over $\reals$.  Translating the absolute value function to 
its piecewise equivalent 
(see \ref{abs}), the results of carrying out the arithmetic as above gives
$$\begin{cases}
x\mapsto 0 & \lambda<0 \\
x\mapsto 0 & \lambda=0 \\
x\mapsto 0 & \lambda>0.
\end{cases}
$$
which is extensionally equivalent to $0$, but not intensionally equal to $0$.
Thus we need a further normalization step which would
combine the above redundancies.  

\begin{defn}
Let $D=(F,\sim, C)$ be a strong effective domain of functions, where $F:O\to V$
and $O$ is a linearly ordered domain.  We call
$\mathbb{P}_D(\fin(|O|),F)$ an \emph{effective piecewise domain}.
\end{defn}

Note that we are not assuming that all the functions in $F$ are total - only that
we have an effective method ($\sim$) for deciding equivalence.  The first
simplification algorithm is then very simple to describe: apply $C$ to each
part of a piecewise function $p$ giving a new function $q$ with the same
breakpoint set, and then merge (in increasing order) adjoining triples
$(g_{2i-1},g_{2i},g_{2i+1})$ if they are all equal.  More precisely, 
Figure~\ref{pseudoalgo} gives Ocaml code to implement this.
\begin{figure}
\begin{algorithm}\label{pseudoalgo}Pseudo normal form
\begin{small}
\begin{verbatim}
type ('a,'b) condpair = 
    {left_fn : ('a -> 'b); 
     pt_fn : ('a -> 'b); right_pt : 'a}
and ('a,'b) endpiece = {fn : ('a -> 'b)}
and ('a,'b) piecewise = 
  (('a,'b) condpair) array * ('a,'b) endpiece ;;

let pseudonormalform (normal:('a->'b) -> ('a->'b)) 
  ((a,e):('a,'b) piecewise) : ('a,'b) piecewise = 
  let pnormal y = {y with left_fn = normal y.left_fn; 
                          pt_fn = normal y.pt_fn}
  and canmerge a b =
    a.left_fn == a.pt_fn && a.pt_fn == b.left_fn
  and merge a b = {left_fn = a.left_fn; 
      pt_fn = b.pt_fn; right_pt = b.right_pt}
  in
    let (b,newe) = (Array.map pnormal a, {fn = normal e.fn})
    and j = ref 0
    and n = Array.length a in 
      if n=0 then (b,newe)
      else begin
        for i=1 to n-1 do
          if canmerge b.(!j) b.(i) then
            b.(!j) <- merge b.(!j) b.(i)
          else
            j := !j + 1;
        done;
        if b.(!j).left_fn==b.(!j).pt_fn && 
          b.(!j).pt_fn==newe.fn then
          (Array.sub b 0 !j, newe)
        else
          (Array.sub b 0 (!j+1), newe) 
        end;;
\end{verbatim}
\end{small}
\end{algorithm}
\end{figure}
\noindent
\emph{normal} is the normalizing function $C$ of $D$.  Note that we have
used a record structure to {\em statically} enforce the fact that any piecewise
function with $n$ breakpoints must consist of $2n+1$ functions, with adjoining
regions alternating between connected open sets and a point upper bound, and
ending with a single (upward) unbounded piece.  It is possible to use a simpler
data-structure for this and simplify the code, but we would 
lose the ability to statically enforce some important invariants.

\begin{prop}\label{preserves}
Let $f\in \mathbb{P} = \mathbb{P}_D(\cfin(|O|),F)$, where $\mathbb{P}$ is an
effective piecewise domain, 
then $\pnform(N,f)$ and $f$ are extensionally equal, where $N$ is any function
$N:F\mapsto F$ which preserves $\simeq$.  Additionally, if $N$ is idempotent, 
then so if $\pnform$.
\end{prop}

A complete proof can be found in Appendix~\ref{AppA}.
Unfortunately, this simple algorithm does not actually give a normal form, 
never mind a canonical form, even
if we restrict our input functions to polynomials over $\ZZ$.  Consider for
example
$$\begin{cases}\label{spurious_disc}
x\mapsto 0 & \lambda<0 \\
x\mapsto x^2 & \lambda=0 \\
x\mapsto 0 & \lambda>0.
\end{cases}
$$
which ``simplifies'' to itself.  Of course, the above function is extensionally
equal to $0$, so we do not in fact have a complete normal form.  However, for
some restricted classes of functions, this does give a normal form.

\begin{prop}
Let $f\in \mathbb{P}_D(\fin(|O|),F)$ be such that for all $g_{2i}$ components
of $f$ defined on the points of the range partition associated with $f$, then
either $g_{2i} \simeq g_{2i-1}$ or $g_{2i} \simeq g_{2i+1}$.  For such $f$, the
\texttt{pseudonormalform} algorithm gives a normal form.
\end{prop}

\noindent
The proof is straightforward.  The proposition can be understood to say that if
the function we are dealing with has a representation into pieces that are
somehow compatible with each other (i.e. applying $C$ is enough to recognize
this), then we have a normal form.  To get a complete normal form, we have to
figure out if, at the breakpoint, the function is ``compatible'' with its
neighbours.  To understand why this is not so simple, consider

$$\begin{cases}
x\mapsto 0 & \lambda<0 \\
x\mapsto \delta_0(x) & \lambda=0 \\
x\mapsto 0 & \lambda>0,
\end{cases}
$$
where $\delta_a(x)$ is the usual characteristic function of the point $a$.
To be able to properly handle such cases, de-nesting of piecewise-defined
functions is necessary.

Consider our first algorithm \ref{pseudoalgo}, but with the \texttt{canmerge}
function defined as
\begin{verbatim}
let canmerge' a b = ((a.left_fn = b.left_fn) && 
  (a.pt_fn a.right_pt == b.left_fn a.right_pt))
\end{verbatim}

In other words, we merge 2 consecutive pieces if and only if the normal forms
for the functions on the two open intervals are the same and if the point
function and left hand function evaluate to the same value.  More precisely,

\begin{algorithm}\label{algo1} Let\\
\texttt{canonform p = pseudonormalform' (denest p)}\\
where \texttt{pseudonormalform'} is obtained from
\texttt{pseudonormalform} by replacing \texttt{canmerge} with \texttt{canmerge'}.
\end{algorithm}

The \verb+denest+ function is a simple linear traversal (specified by example
in subsection~\ref{denest}) which brings (definitionally) nested piecewise 
functions to the surface.  This does not increase the {\em total} number
of breakpoints, although it usually increases the number of outer breakpoints.

\begin{theorem}\label{main_thm}
Let $f\in \mathbb{P}=\mathbb{P}_D(\cfin(|O|),F)$, where $\mathbb{P}$ is an
effective piecewise domain, and $f$ is such that for all breakpoints
$b\in\cfin(|O|)$, there exists a decision oracle $=_V$ for equality of values.
In other words, for all $g_1,g_2\in F$ and all $b\in\cfin(|O|)$, it is possible
to decide if $g_1(b) = g_2(b)$ with $=_V$.  Then algorithm \ref{algo1}
is a canonical form algorithm.
\end{theorem}

The full proof is in Appendix~\ref{AppA}.
While the above may appear to give a qualified normal form, it
nevertheless turns out to be extremely useful in practice, as very wide classes
of examples are covered.  Instead of using a function $=_V$ on values,
one instead uses a semi-decision procedure
for $\neq_V$, and only structural equivalence for $=_V$. 
While this no longer gives a normal (or canonical) form, for many practical
examples this appears to be sufficient.

\begin{cor}
Let $O=\reals$, restrict $\cfin(|O|)$ to the real
algebraic numbers, and $F$ to be rational functions, then algorithm
\ref{algo1} gives a canonical form algorithm.
\end{cor}

\section{Complexity}\label{complexity}

We are primarily interested in comparing complexity results between our
approach and that of \cite{Mohrenschildt}, and thus we will restrict ourselves
to a setting where this comparison can (fairly) be made.  Although we would
have preferred to make this paper self-contained, repeating all the necessary
definitions from \cite{Mohrenschildt} would take us too far afield, and we will
be forced to assume that the reader has a certain familiarity with its
contents.

Without loss of generality, we can assume that arithmetic operations on 
function \emph{representations} are $O(1)$, and that the normal form operation
$C$ on function representations is $O(M(n))$ where
$n$ is the size of the representation.  It is then easy to obtain that

\begin{prop}\label{OurCompl}
Let $f$ be a piecewise function (as per Theorem~\ref{main_thm})
with $d$ breakpoints, with each $g_i$ bounded
in size by $n$.  Then Algorithm \ref{algo1} runs in $O(d M(n))$.
\end{prop}

Naturally for complex expressions, $M(n)$ can still be the driving factor in
the overall cost.  The cost above is trivial to establish as Algorithm~\ref{pseudoalgo}
only does at most $4$ linear traversals of the expression (once for denesting,
the \texttt{Array.map}, the middle loop, and the final \texttt{Array.sub}).  Only
the middle loop needs to perform non-trivial computations.

\begin{prop}\label{MvMCompl}
Under the same assumptions, von Mohrenschildt's algorithm~\cite{Mohrenschildt} 
runs in $O(2^d M(n))$.
\end{prop}
The reason for this is that the algorithm of section 6.1 of~\cite{Mohrenschildt}
expands piecewise expressions into terms which the normal form algorithm steps
(3.4) and (3.6) (section 3) further expand.

\section{Remarks}

For lack of space, we did not include here the full algorithm for 
definitional denesting.  However this is quite straightforward.
Denesting of composed piecewise functions is considerably more difficult;
however, the key ideas are in von Mohrenschildt's work~\cite{Mohrenschildt},
and these can be combined with our the ones in the present work.  The
principal difficulty here remains that of ``inverting'' functions to
create a \emph{finite} set of breakpoints.  This is why~\cite{Mohrenschildt}
restricts the inner functions to be polynomials.

\section{Conclusions and Further work}\label{future}

In the current work, we make the following contributions: a simple yet
general exposition of piecewise functions that cleanly separates the 
decision aspects from the value aspects of these functions; this allows us
to leverage the underlying linear structure to give faster algorithms
(linear instead of exponential in the number of breakpoints); a clean
separation of concerns between the requirements on the domain and the
range of piecewise functions; and a clearer picture of the kinds of 
normal and canonical forms needed from the base domains to be able to 
build piecewise functions.

While all our examples are over the $\reals$, it is clear that our work also
applies to finite domains (which can be linearly ordered).  Finite
unions of linearly ordered domains can also be handled - one can just pick 
an arbitrary order between the domains, where none of the domains ``touch''; we
can combine the decision procedure $=_{V}$ for each of the sub-domains to a
decision procedure for the full domain.

For example, by using a logic which can deal with partial functions and
undefinedness~\cite{Farmer90,Farmer93b}, the functions we deal with can be
partial.  This was our original motivation for looking into this problem!
The issue with von Mohrenschildt's work is that it needs a {\em ring} in
both the value and range domains.  Here, we only require sets with operations
and a normalization procedure in the range, and ordering properties in the
domain.

In the future, we hope to move from linearly ordered domains to domains with
finite presentations and algorithmic combination properties.  The main
examples, of course, being algebraic and semi-algebraic sets in $\reals^n$, where 
respectively Gr\"{o}bner Bases and CAD are the algorithmic tools.  Another
source of generalization might be to work with implicit characteristic functions,
so as to be able to handle functions like \texttt{floor} and \texttt{trunc}.

\bibliographystyle{abbrv}
\bibliography{../allcite,../jacques}

\appendix
\section{Proofs}\label{AppA}
\begin{prop2}[\ref{preserves}]
Let $f\in \mathbb{P} = \mathbb{P}_D(\cfin(|O|),F)$, where $\mathbb{P}$ is an
effective piecewise domain, 
then $\pnform(N,f)$ and $f$ are extensionally equal, where $N$ is any function
$N:F\mapsto F$ which preserves $\simeq$.  Additionally, if $N$ is idempotent, 
then so if $\pnform$.
\end{prop2}

\begin{proof}
Let $f$ have the following form:
$$
\begin{cases}
g_1(x) & x\in\Lambda_1\\
g_2(x) & x=\lambda_1\\
\vdots & \vdots\\
g_{2n}(x) & x=\lambda_n\\
g_{2n+1}(x) & x\in\Lambda_{n+1}\\
\end{cases}
$$
We will work on the abstract mathematical form as above, since the mapping
between it and the data-structure \texttt{('a,'b) piecewise}
is obvious, but the mathematical notation is clearer.
Remember that the decomposition above satisfies the following pre-condition,
$$ \forall x\in\Lambda_i, x\leq \lambda_i \wedge 
   \forall 1\leq j < n, \lambda_j < \lambda_{j+1} $$
and that the $\lambda_i$'s form a range partition for $\Lambda$.

The first step of the algorithm transforms this to a function $f_1$ (stored in
$(b,newe)$ in the code) of the same form but with $g_i$ replaced with
$\tilde{g}_i := N(g_i)$; however, by definition, $N$ preserves $\simeq$, and
thus $f_1\simeq f$.  

We then have 3 different cases: $n=0$, $n=1$, and $n>1$.  For $n=0$, then our
piecewise function is already in one piece, no reduction can be done, and
so the normalized form can be returned.

For $n=1$, the \texttt{for} loop is empty.  The \texttt{if} condition is a
specialized version of \texttt{canmerge} for comparing a general piece to the
end piece.  The condition amounts to checking that all 3 functions involved are
the same, in which case we can simply make the range partition null, which is
what is done via \texttt{Array.sub b 0 \!j} (the sub-array of b starting at
index 0 of length $j$) as $j=0$.

For $n>1$, the \texttt{for} loop actually encodes a fold function in an
imperative manner.  It is easy to see that, for any $i$, $j\leq i$ and that the
function defined by the first $i$ pieces of $f$ and the one contained in the
$j$ pieces of $b$ are extensionally equivalent (by construction and the
properties of \texttt{canmerge} and \texttt{merge}).  This is the invariant
maintained by the loop.  On exit of the loop, we perform a last step on the
last general piece and the end piece, as already analyzed above in the $n=1$
case.

To prove idempotency, assuming that $N$ is idempotent, the same proof skeleton
as above works.  The only difference is in the case of $n>1$, where we must
show that a second (linear) pass through the pieces of $f_1$ would not merge
any more segments.  But if a second pass were to merge two pieces, we can show
that these pieces would have been merged already in the first pass.
\end{proof}

\begin{theorem2}[\ref{main_thm}]
Let $f\in \mathbb{P}=\mathbb{P}_D(\cfin(|O|),F)$, where $\mathbb{P}$ is an
effective piecewise domain, and $f$ is such that for all breakpoints
$b\in\cfin(|O|)$, there exists a decision oracle $=_V$ for equality of values.
In other words, for all $g_1,g_2\in F$ and all $b\in\cfin(|O|)$, it is possible
to decide if $g_1(b) = g_2(b)$ with $=_V$.  Then algorithm \ref{algo1}
is a canonical form algorithm.
\end{theorem2}

\begin{proof}
First, we can re-use the proof of \ref{preserves}, but with the 
modified \texttt{canmerge'} to show that extensional equivalence is
maintained.  Since
\verb+a.pt_fn = b.left_fn+ implies 
\verb+a.pt_fn a.right_pt == b.left_fn a.right_pt+, the new
algorithm will definitely merge all previously merged pieces, and may
merge more.  But, similarly to the previous proof, extensional equivalence
is always maintained.  In fact, the modified \texttt{canmerge'}
checks this quite explicitly for the breakpoints.  Actually, using the same
proof, since $C$ preserves decidable equivalence, we have the stronger result
that $\sim$ is preserved.

We have to prove that if $f_1$ and $f_2$ are two functions in
$\mathbb{P}_D(\cfin(|O|),F)$ then $\cform(f_1) = \cform(f_2)$ if and only if
$f_1\sim f_2$.

Without loss of generality, we can assume that neither $f_1$ nor $f_2$ are
nested.

$\Rightarrow$.  Suppose $\cform(f_1) = \cform(f_2) = \tilde{f}$.  But we have
already shown that $\cform(f_1)\sim f_1$ and $\cform(f_2)\sim f_2$, and as
$\sim$ is symmetric, this implies that $f_1\sim f_2$.

$\Leftarrow$.  Suppose $f_1 \sim f_2$.  But $\cform(f_1)\sim f_1$ and
$\cform(f_2)\sim f_2$, and by symmetry, $\cform(f_1)\sim\cform(f_2)$.
However, we need to prove that these are in fact equal.  Suppose that they are
not.  They either they have a different set of breakpoints or (assuming that they
have the same breakpoints), that the underlying functions differ.  Let
$k_1 = \cform(f_1)$ and $k_2 = \cform(f_2)$.

First, suppose that the set of breakpoints is different.  Without loss of generality,
assume that it is $k_1$ and is in fact the first breakpoint $\lambda_1$.  Let 
$k_1$ have the
same form as $f$ in the preceding proof (with sub-functions labelled $g_i$), and 
use $h_j$ for the labels of the sub-functions of $k_2$.  Now the $g_i$'s and the
$h_i$'s are breakpoint-free functions for which we have a canonical form $C$
(by assumption).  Since $g_1\sim h_1$, $C(g_1) = C(h_1)$; 
$g_2(\lambda_1) =_{V} h_1(\lambda_1)$ (since $f_1\sim f_2$); and
$g_3\sim h_1$ implies $C(g_3) = C(h_1) = C(g_1)$.  But since $C(g_1) = C(g_3)$
and $h_1$ is defined at $\lambda_1$, so is $g_1$.  By extensionality and the
fact that $h_1$ is breakpoint free, \texttt{canmerge'} applied to the first
breakpoint of $k_1$ would have merged this part -- contradiction.

Second, suppose that the breakpoints are the same, but the sub-functions are different.
This is not possible either because $C$ is, by definition, a canonical form for the
underlying sub-functions.
\end{proof}

\end{document}